\documentstyle [psfig]{elsart}
\begin{document}
\begin{frontmatter}

\title{Step Patterns on Vicinal Reconstructed Surfaces}
\author{Igor Vilfan}
\address{J. Stefan Institute, P.O. Box 100, SI-61111 Ljubljana, Slovenia\\
e-mail: igor.vilfan@ijs.si}

\begin{abstract}
Step patterns on vicinal $(2\times1)$ reconstructed surfaces of noble 
metals Au(110) and Pt(110), miscut towards the (100) orientation, 
are investigated. 
The free energy of the reconstructed surface with a network of crossing 
opposite steps is calculated in the strong 
chirality regime when the steps cannot make overhangs. 
It is explained why the steps are not perpendicular to the direction 
of the miscut but form in equilibrium a network of crossing steps which
make the surface to look like a fish skin.
The network formation is the consequence of 
competition between the -- predominantly elastic -- energy loss and 
entropy gain. It is in agreement 
with recent scanning-tunnelling-microscopy observations on 
vicinal Au(110) and Pt(110) surfaces. 
\end{abstract}

\end{frontmatter}

\section{Introduction}

Vicinal surfaces inevitably have steps which are, in general, perpendicular
to the miscut direction. 
For the $(2\times 1)$ reconstructed surfaces 
like Au(110) and Pt(110), however, the situation is different 
when the surface is miscut towards the (010) orientation.
Recent scanning-tunnelling microscopy (STM) 
observations revealed that the single-height difference steps 
are not parallel to $[00\bar1]$, but run
roughly in the $[1\bar11]$ and $[1\bar1\bar1]$ directions \cite{GCBE91,GBS92}.
The steps thus cross and form a network, also called  "textured pattern"
or "fish skin," according to
the specific shape of the terraces bound by the steps.

\begin{figure}
 \centerline{\psfig{file=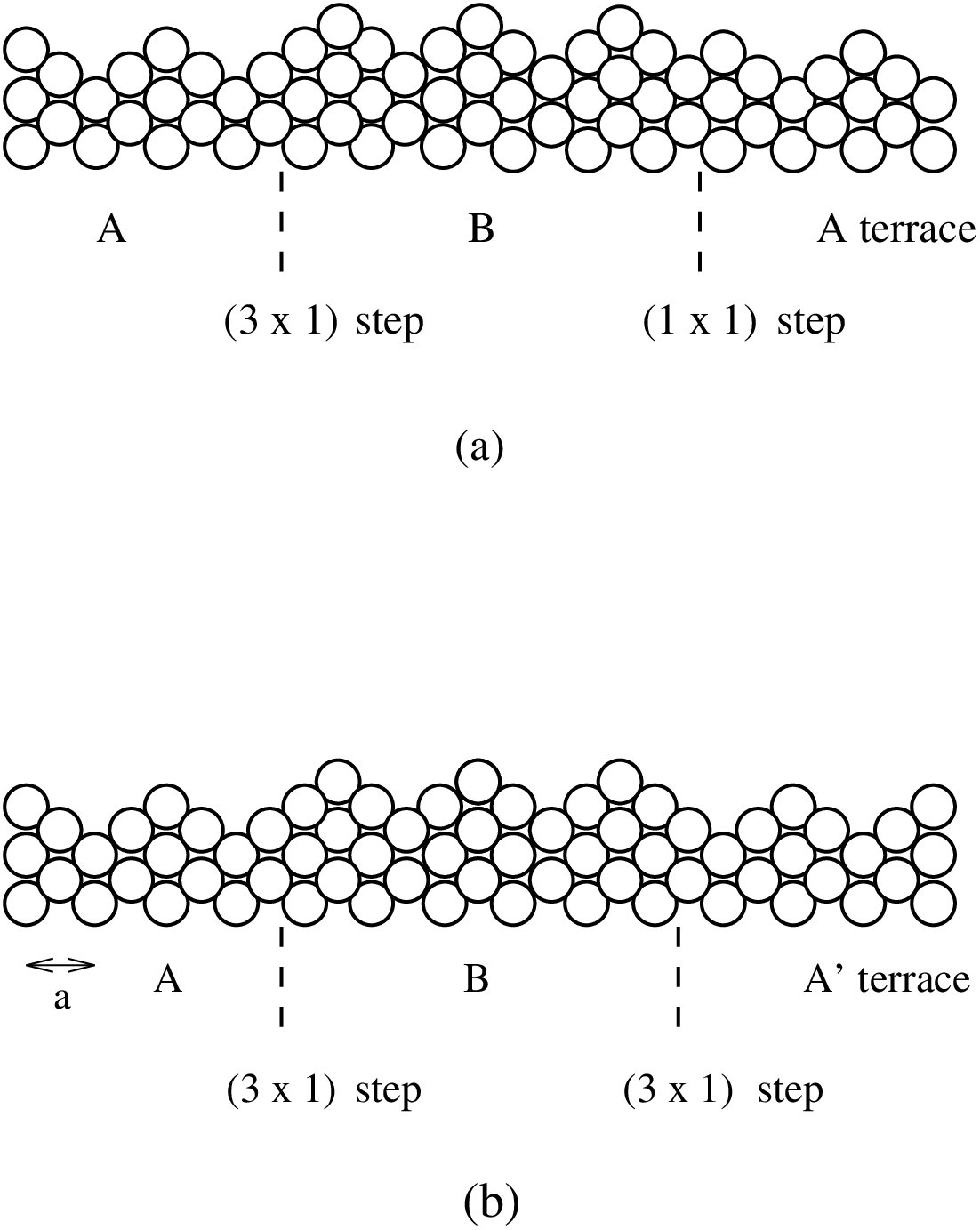,width=9truecm,angle=0}}
 \caption{Cross section through the $(2\times 1)$ reconstructed (110) surface 
 of an FCC crystal.
 (a) A $(3\times 1)$ and a $(1\times 1)$ step. 
 (b) Two opposite $(3\times 1)$ steps.}
 \label{steps}
\end{figure}

\begin{figure}
 \centerline{\psfig{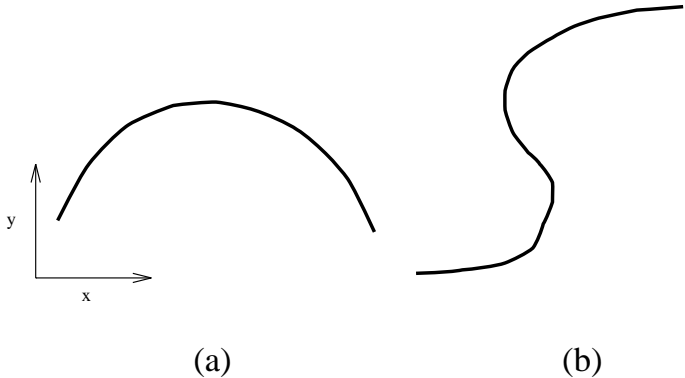}}
 \caption{Overhangs of a step on a reconstructed surface. (a) Because of the
 $(2\times 1)$ reconstruction, the overhangs
 in the $y$ direction 
 of a $(3\times 1)$ step alone 
 are not allowed, only overhangs where the $(3\times 1)$ step
 changes into a $(1\times 1)$ are possible. (b) The overhangs in the
 $x$ direction increase the length, projected on the
 $x$ axis and thus increase the number of broken atomic bonds.}
 \label{overhangs}
\end{figure}

A specific property of the steps on reconstructed FCC (110) surfaces is 
the anisotropy in their energy.  
The step energy is high (of the order $0.1$ eV) for steps perpendicular to the 
missing rows and low ($\sim 10^{-3}$ eV)
when they are parallel to the missing rows.
Two types of low-energy single-height-difference steps can be formed, 
the $(3\times1)$ and the $(1\times1)$ steps, after the size of their 
exposed $(111)$ facets, see Fig.\ \ref{steps}. 
The two types of steps have opposite  
lateral shifts in the topmost atomic row positions 
so that the rows on the  right hand and on the left  hand 
terraces are in phase. In general, the two types of steps have different 
energies. Calculation with the embedded-atom method tells us that 
the $(3\times1)$ steps have lower energy than the $(1\times1)$ steps
\cite{RFDB90}.
Indeed, the $(3\times 1)$ are also the predominant steps seen in STM on Au(110)
and Pt(110) \cite{GCBE91,GBS92}.  
The difference in energy is connected with chirality, i.e., the $A-B$ and
the $B-A$ steps are not equivalent \cite{Fish86}. In the strong chirality 
regime when the energy of the $(1\times 1)$ step is much higher than the energy
of the $(3\times 1)$ step, only the $(3\times1)$ steps are thermally
excited.
Here we will consider the strong chirality case so that the 
presence of the $(1\times1)$ steps is negligible. 
At each $(3\times1)$ step the phase in the position of the topmost
atomic rows is shifted laterally by $+\pi/2$. 
Two $(3\times1)$ steps together thus cause a phase shift of $\pi$, they 
form an Ising (antiphase) domain boundary, the missing rows on the 
terraces $A$ and $A^\prime$ 
are out of phase, see  Fig. \ref{steps}(b). 
For this reason, the $(3\times 1)$ steps cannot make 
overhangs (loops back, shown in Fig. \ref{overhangs}(a)) 
in the $[\bar110]$ direction 
\cite{VV90}. If the overhangs were allowed, then one can get from $A$ to
$A^\prime$ around the overhang without crossing any domain boundary or step,
so the $A$ and $A^\prime$ terraces should be in phase. 
Like in the four-state Potts model, (a multiple of) four steps must meet 
in a vertex (at least two lines cross or four lines terminate).

\begin{figure}
 \centerline{\psfig{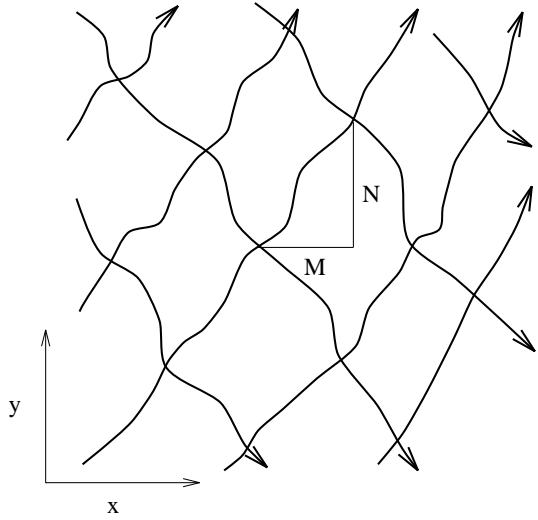}}
 \caption{View on a vicinal reconstructed (110) surface, miscut towards (100).
 The missing rows are parallel to the $y$ axis.
 The $(3\times 1)$ steps are either "up" or "down" and form a 
 characteristic pattern. Terraces between the steps 
 resemble on a fish skin. The steps make no
 overhangs and the steps of the same kind never cross. The average separations
 between the step crossings, vertices, are $M$ and $N$ in the $x$ and $y$
 directions, respectively.}
 \label{patterns}
\end{figure}

As already mentioned, a miscut of a $(110)$ surface towards the
$(010)$ orientation is expected to 
produce steps, parallel 
to the $[00\bar1]$ ($x$) axis. The energy of such steps is very high
because some additional surface-atoms bonds are broken there.
Meandering of steps in the $y$ ($[\bar110]$) direction, on the other hand, 
creates sections, parallel to the $y$ direction, which have much lower energy 
than the sections parallel to the $x$ axis.
At finite temperature, therefore, the steps are not straight lines along the $x$
axis but make many thermally excited excursions in the $y$ direction.
However, as the overhangs in the $y$ direction are forbidden 
in the strong chirality regime,
and the overhangs in the $x$ direction are 
energetically costly and rare as they increase the number of broken bonds
(see Fig. \ref{overhangs}(b)), 
the step excursions are possible only in one direction. 
Depending on the orientation of steps, 
the step excursions are either in the $+y$ 
("up," left-hand step in Fig. \ref{steps}(b)) or in the $-y$ direction 
("down," right-hand step in Fig. \ref{steps}(b)).
Notice that the steps of the same kind cannot cross.
When the imposed inclination is exactly in the $y$ direction, there must be 
an equal
density of "up" and "down" steps which inevitably interpenetrate, the steps 
form a network, shown schematically in Fig. \ref{patterns}, which was also observed in the 
scanning-tunnelling microscopy \cite{GCBE91,GBS92}. 

The aim of this paper is to investigate 
the stability of step patterns. Special emphasis will be on
thermally-induced but chirality-restricted meandering of steps and on the 
elastic interactions between them.
In the following Section we shall analyse the energy and entropy 
of different step configurations and of the step crossings. The elastic 
interaction energy between the steps, calculated in Section 3, will
be used to evaluate the free energy in Section 4.  
In Section 5 the model will be 
applied to the reconstructed surfaces of Au(110) and Pt(110).

\section{Step Energetics}

At a single-height step which is perpendicular to the missing rows,  
one atomic bond along the missing rows is broken per unit length $a$. 
The step energy $V$ (per unit length) is high ($V \gg k_BT$), a step
is seldomly thermally excited in this direction. 
However, with the miscut towards the $(010)$ orientation, 
such steps are imposed on the surface. 
The average step separation in the $y$ direction (measured in 
units of atom spacings $a/\sqrt{2}$) is
$N + {1/2} = (2\tan\theta)^{-1}$ where $N$ is the average number of atoms 
in a top row 
on a terrace and $\theta$ the miscut angle. 
Meandering without overhangs introduces sections of steps parallel to the $y$ 
direction but maintains the total number of broken bonds constant. 
The energy of such sections (per $a/\sqrt{2}$) is denoted by 
$\epsilon$ and is  much smaller than $V$. 
If we neglect the energies of kinks and  take into account that 
the overhangs are forbidden by the strong chirality condition, 
the energy of a step that extends $M$ unit 
cells in the $x$ and $N$ atom spacings in the $y$ directions 
(see Fig. \ref{patterns}) is simply $M V + N \epsilon$. 

Meandering of steps introduces also vertices, 
crossings of "up" with "down" steps. 
At a vertex two opposite steps meet and form a short segment 
(length $N^\prime\ll N$) of a narrow Ising 
domain wall oriented parallel to the 
$y$ axis and with the energy $\eta N^\prime$ \cite{VV91}. 
The vertex energy is $W = (\eta - 2\epsilon) N^\prime $ and it is related to the 
short-range elastic interaction between the steps.  
From the experimental evidence of separate Ising (deconstruction) 
and roughening (or wetting) transitions \cite{CMTP,KZR+94}, the conclusion was
reached that the interaction is weak but attractive, i.e.,
$\eta < 2 \epsilon$.  
The average density of vertices is $(NM)^{-1}$. 
$N$ is determined by the miscut angle and $M$ will be a variational parameter, 
related to the average angle $\phi$ at which two opposite steps cross:
$\tan (\phi/2) = \sqrt{2}M/N$.

Thermally induced step fluctuations increase the step energy on one hand 
but also the entropy on the other hand. 
In the following, we shall approximate the step texture by a regular lattice of 
vertices connected by step segments which fluctuate without overhangs.
A step segment with fixed ends at $(0,0)$ and $(M,N)$ has 
\begin{equation} 
P = {(M+N)! \over M! N!}
\end{equation}
different configurations with the same step energy. The free energy of such a
segment and one vertex is 
\begin{equation}
F(M,N) = M V + N \epsilon -k_B T \ln {(M+N)! \over M! N!} + W.
\end{equation}
This free energy expression is minimal at finite $M/N$. However, this 
step free energy is not complete and the minimum is at wrong 
$M/N$. In the following Section the elastic contribution to the free energy will
be calculated.

\section{Elastic Interaction Between Steps}

Long-range elastic interaction between steps has been first discussed by
Marchenko and Parshin \cite{MP80}; see also \cite{Noz92,Lap94,SPG94}. 
A step exerts elastic forces which result in an elastic 
displacement field in the bulk. 
The displacement field of a step interacts with the elastic forces of other
steps and the interaction energy is the elastic contribution to 
the surface free energy density.

In the continuum elasticity theory, the force on the
underlying elastic continuum exerted by the step $i$ running
parallel to the $y^\prime$ axis (which is, in general, not
parallel to $y$) 
is described by a force doublet localized at the step $i$ 
\cite{MP80,Noz92,Lap94}:
\begin{equation}
   \vec{F}_i(x^\prime) = \vec{f}\delta^\prime(x^\prime).      
   \label{Force}
\end{equation}
$f_{x^\prime} = f_\parallel$ describes the local stretch of the surface which 
is the same for the "up" and "down" steps, and 
$f_z = \tau a/\sqrt{8}$ is a local torque, which tends to twist
the crystal around the $y^\prime$ axis, $\tau$ is the (tangential) 
surface stress.

As usually, we will calculate the elastic energy neglecting the 
fluctuations of the step (assuming straight lines) and assuming an
isotropic bulk.

\subsection{Elastic energy of parallel steps}

The elastic interaction energy between two parallel steps 
$i$ and $j$ a distance $\ell$ apart is \cite{MP80,Noz92,Lap94}:
\begin{eqnarray}
    E_{ij} = {2(1-\sigma^2)\over \pi E}{\vec{f}_1\cdot\vec{f}_2\over \ell^2}
    = {2(1-\sigma^2)\over \pi E}{f_\parallel^2 + f_z^2\over \ell^2},
\end{eqnarray}
where $E$ is the Young modulus and $\sigma$ the Poisson ratio. 
When the surface is covered with an infinite array of equidistant 
parallel steps, the step interacts with all the other steps. 
If $\ell$ is the distance between two neighbouring steps, 
the elastic contribution to the surface energy density is
\begin{equation}
    e_p = {1\over \ell}\sum_{i>j}E_{ij} = 
    {\pi(1-\sigma^2)\over 3 E}{f_\parallel^2 + f_z^2\over \ell^3},
\end{equation}
parallel steps of the same kind always repel.

\subsection{Elastic energy of two crossing steps}

Now we will calculate the elastic interaction energy of an "up" step 
with a "down" step that cross at $(x^\prime, y^\prime)=0$. 
Let the step 1 be along the $y^\prime$ axis and let the angle between the 
steps be $\phi$. 
The elastic interaction energy, if both steps are infinitely long, is:
\begin{equation}
    E_{12}(\phi) = - \int_{-\infty}^{+\infty} \d x^\prime \d y^\prime 
    \vec{F}_1(x^\prime)\cdot \vec{u}_2(x^\prime,y^\prime),
    \label{Ephi}
\end{equation}
where $\vec{u}$ is
the displacement field of the second step at $(x^\prime,y^\prime)$, 
\begin{equation}
   \vec{u}_2 = - {2(1-\sigma^2)\over \pi E} {\vec{f}_2\over r}
\end{equation}
and $r$ the shortest distance between the point $(x^\prime,y^\prime)$ and 
the step 2, 
\begin{equation}
    r = \vert x^\prime \cos \phi + y^\prime \sin \phi \vert.
\end{equation}    
Since the two crossing steps are always opposite to one another, 
we have $f_{2,z} = - f_{1,z}$,
whereas $f_{2,x^\prime} = f_{\parallel}\cos \phi$.
Partial integration of (\ref{Ephi}) over $x^\prime$ yields:
\begin{equation}
    E_{12}(\phi)  = 
    {2(1-\sigma^2)\over \pi E}\int_{-\infty}^{+\infty}{\d y^\prime\over r^2}
    [f_{\parallel}^2 \cos^2 \phi + f_{z}^2 \cos \phi].
\end{equation}
Obviously, the continuum theory breaks
down for small $r$, therefore we integrate over $\vert r\vert > a $ and 
incorporate the remaining short-range interaction into the vertex energy $W$.
The energy of two crossing steps is then:
\begin{equation}
    E_{12}(\phi) = {4(1-\sigma^2)\over \pi E a}
     [f_{\parallel}^2  \cos^2 \phi - f_{z}^2 \cos \phi]
    {1\over \sin \phi} + W.
\label{e12}
\end{equation}
The expression (\ref{e12}) diverges for $\phi=0$ when the 
steps overlap. Nevertheless, we notice that the square bracket reduces to
$f_{\parallel}^2 + f_{z}^2$ for parallel steps of the same kind 
($\phi\to \pi$) and to 
$f_{\parallel}^2 - f_{z}^2$ for opposite (antiparallel,
$\phi\to 0$) steps, in agreement with 
\cite{MP80}. From Eq. (\ref{e12}) it is also evident that two 
perpendicular steps do not interact elastically.
 
\section{The free energy density}

Using the results of the previous Sections, the free energy density 
(per unreconstructed surface unit cell) becomes equal to:
\begin{eqnarray}
   f(\xi) =  a + {b\over \xi} + {(c - d \xi^2 )(1-2\xi^2)\over 
   \xi^2(1 + 2\xi^2)}
   + e {(1 + 2\xi^2)^{3/2}\over \xi^3} \cr\cr
   - {k_B T \over N} [\ln ({1+\xi\over \xi}) + {\ln (1 + \xi)\over \xi}]
   \label{fe2}
\end{eqnarray}
where
\[    \xi = {M\over N}  \]

\[
    a={V\over N},       \]
\[  b={1\over N}(\epsilon + {W\over N}),     \]

\[  c={\sqrt{2}(1-\sigma^2)\over \pi E N^2a}(f_{\parallel}^2 - f_{z}^2),
\]

\[  d={2\sqrt{2}(1-\sigma^2)\over \pi E N^2a}(f_{\parallel}^2 + f_{z}^2),
\]

\[  e={\pi\sqrt{2}(1-\sigma^2)\over 24 E N^3 a}(f_\parallel^2 + f_z^2)
    = {\pi^2\over 48 N} d.                   \]
 
The first term in (\ref{fe2}) is the energy of a step, parallel to
the $x$ axis. At constant miscut angle, the first term  does not influence the 
equilibrium shape of the step patterns, it only shifts the energy. 
The second term is the contribution of the step segments 
that are parallel to the $y$ axis and a contribution from the vertex energy. 
The third and the fourth terms are the elastic energies of crossing and parallel
steps, respectively. 
The last term is the entropy contribution to the free energy assuming a  
regular lattice of vertices.
The entropy tends to increase the density of steps and to decrease  
the angle between the opposite steps. 
Since the entropy is calculated for a step fluctuating only within
a finite, $M\times N$ rectangle,  the "loss of entropy"
which leads to a $\sim 1/\ell^2$
repulsion between parallel steps \cite{Noz92} is properly taken into account 
in (\ref{fe2}).
   
In principle, the ratio $\xi = M/N$ is arbitrary, $0 < \xi < \infty$.
However, the free energy diverges for $\xi\ll 1$ 
and goes to $(d + 2\sqrt{2} e)$ for $\xi\gg 1$.
$\xi$ is determined from the condition that $f$ is minimal and it is finite
at finite temperature.

\section{The case of A\lowercase{u}(110) and P\lowercase{t}(110)}

The above results will now be applied to investigate the stability of 
step patterns on Au(110) and Pt(110) surfaces
at room temperature ($k_B T = 25$ meV). At this temperature,
the surface diffusion (of Au) is probably high enough to enable equilibration of steps on 
lengthscales of $\approx 100$ {\AA} in less than an hour or so 
\cite{SMRBH94}. 
The parameters, used in the estimates are collected in Table \ref{parameters}.
Since $f_\parallel$ is not known, we take $|f_\parallel/f_z| = r$ as a
parameter. 
The vertex energy is small ($W/N \ll \epsilon$), therefore it will be neglected
in calculating the free energy ($W=0$).

\begin{table}[hb]
\caption{Parameters of the model for Au and Pt.} 
\label{parameters}
\begin{tabular}[]{ c  c  c  c  c } 
        \\ \hline\hline
        & $\epsilon$   &  $E$      & $\sigma$ & $\tau$  \\ 
        & $10^{-3}$ eV &  GPa      &          & J/m$^2$ \\ \hline
Au(110) &  1.8 $^a$    &  80 $^b$   & 0.42 $^b$ & 1.0 $^c$ \\
Pt(110) &  2.3 $^d$    &  147 $^b$  & 0.39 $^b$ & 1.6 $^c$ \\ \hline\hline
\end{tabular}\\
$^a$From Ref.\ \cite{RFDB90},\\
$^b$Annealed metal, from Refs.\ \cite{SPG94} and \cite{AIP} ,\\
$^c$Tangential surface stress of spherical crystallites, 
    from Ref.\ \cite{SNS94},\\
$^d$Estimated on the basis of the critical temperatures for 
    Au(110) and Pt(110).
\end{table}

\begin{figure}
 \centerline{\psfig{file=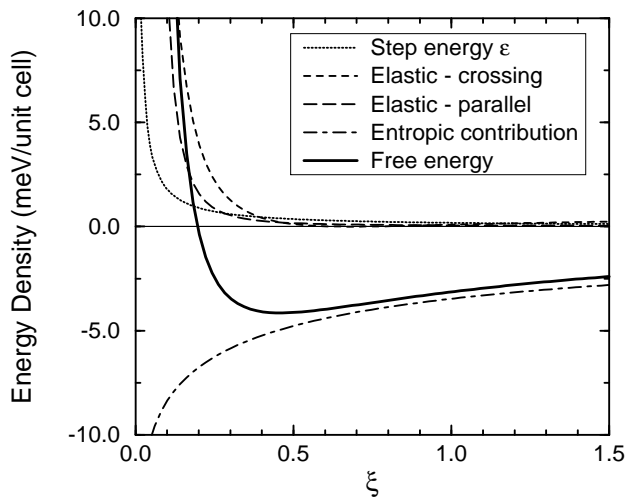,width=9truecm,angle=0}}
 \caption{Free energy density $f$ and individual contributions to $f$ for 
Au(110) surface with $2.7^\circ$ miscut ($N=10$), $W=0$ and $r=4$. 
The free energy density has a minimum at $\xi=M/N\approx 0.4$.}
 \label{free_energy}
\end{figure}

Fig. \ref{free_energy} shows individual contributions to the free energy 
density (\ref{fe2}) for vicinal Au(110) surfaces with $W=0$ and $r=4$.
The free energy density is minimal at $\xi\approx 4$. 
At large $\xi$ (when the steps are almost parallel to the $x$ axis), the
entropy term prevails whereas at small $\xi$ (when very dense steps are almost 
perpendicular to the $x$ axis) the energy density prevails. At short $\xi$, 
the elastic interaction energy is more important than the step energy.
For $r>\sim 2$, the dominant contribution to the energy density comes 
from elastic interactions between crossing opposite steps. 

\begin{figure}
 \vskip5truemm
 \leftline{\psfig{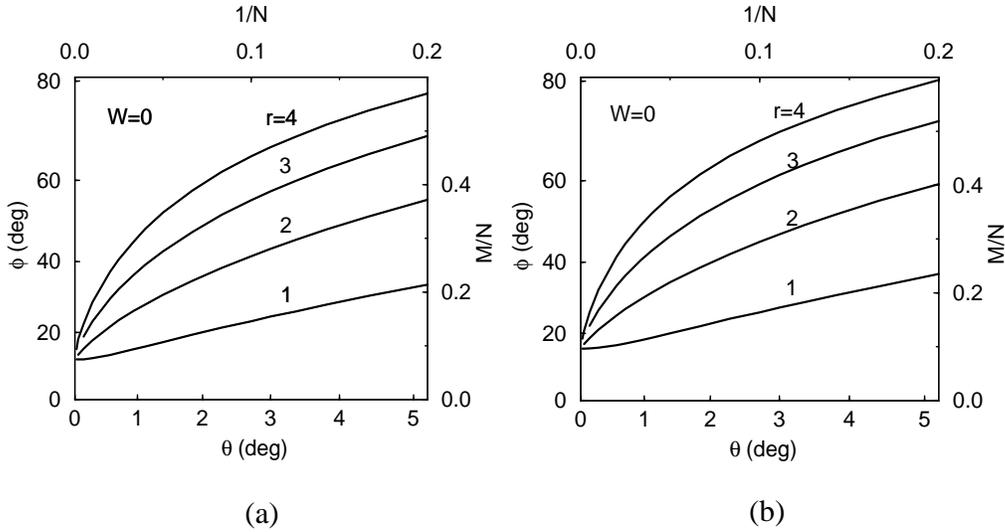}}
 \caption{Average angle between crossing steps, $\phi$, as a function of the 
 miscut angle $\theta$ for various values of $r=|f_\parallel/f_z|$ and $W=0$. 
 (a) Au(110), and (b) Pt(110).}
 \label{phi}
\end{figure}

The equilibrium values of the average angle $\phi$ between opposite steps 
as a function of the miscut angle $\theta$  for  reconstructed Au and Pt (110) 
surfaces are shown in Fig.\ \ref{phi}. Even for vanishingly small miscut angles
the steps form the specific patterns. 
In the limit as $\theta\to 0$, when  
the step separation diverges ($N\to \infty$), the elastic energy 
contributions per surface unit cell vanish as $1/N^2$ or as $1/N^3$, 
and the equilibrium step-crossing angle is given by
\begin{equation}
   \phi = 2 \tan^{-1}[\sqrt{2}(\mathrm{e}^{\epsilon/k_BT}-1)].
\end{equation}
Only in the limit as $T\to 0$ the steps become perpendicular to the miscut
direction -- provided they are not frozen in a finite-temperature configuration.

\section{Conclusions}

It was shown that thermally activated fluctuations of steps, 
constrained by the conditions of 
no overhangs and no crossings of the steps of the same kind, lead to the
characteristic step patterns. 
For the missing-row reconstructed surfaces, the overhangs in the 
direction of missing rows are forbidden or appear with very small probability
in the strong chirality regime and 
the overhangs in the direction perpendicular to the missing rows are 
virtually impossible  if the anisotropy in the step energy is strong enough
($V \gg \epsilon$). 
Both conditions seem to be fulfilled in the case of the reconstructed (110) 
noble metal surfaces, like Au(110) or Pt(110).
The angle under which the opposite steps cross depends on the miscut.
For small miscut angles ($\theta <\sim 0.2^\circ$) or $r<\sim 1.5$,
the patterns are 
determined by competition between the step energy and entropy. 
For $\theta >\sim 0.2^\circ$ and $r >\sim 2$, the competition between the 
elastic interaction energy and the entropy controls the step patterns.
The vertex energy $W$ bears short-range elastic interactions, it 
is usually small and was neglected in this paper.  

It would be interesting to check the temperature dependence of the step 
patterns experimentally. 
By measuring the angle between the steps, one can get 
insight into the elastic interaction energy between the steps. In
particular, one could determine the ratio between the local stretch and
local torque at the steps, $r=|f_\parallel/f_z|$. From the length $N^\prime$
of the sections with two opposite steps running next to each other, 
information on the vertex energy $W$ could also be obtained.  

\ack{The author is indebted to Enrico Carlon who first pointed out the
problem of step pattern formation.}


\begin{thebibliography}{10}

\bibitem{GCBE91}
T. Gritsch, D. Coulman, R.~J. Behm, and G. Ertl, Surf. Sci. {\bf 257},  297
  (1991).

\bibitem{GBS92}
J.~K. Gimzewski, R. Berndt, and R.~R. Schlittler, Phys. Rev. B {\bf 45},  6844
  (1992).

\bibitem{RFDB90}
L.~D. Roelofs, S.~M. Foiles, M.~S. Daw, and M.~J. Baskes, Surf. Sci. {\bf 234},
   63  (1990).

\bibitem{Fish86}
M.~E. Fisher, J. Chem. Soc., Faraday Trans. {\bf 82},  1569  (1986).

\bibitem{VV90}
J. Villain and I. Vilfan, Europhys. Lett. {\bf 12},  523  (1990).

\bibitem{VV91}
I. Vilfan and J. Villain, Surf. Sci. {\bf 257},  368  (1991).

\bibitem{CMTP}
D. Cvetko {\it et~al.}, Surface Science {\bf 269},  68  (1992).

\bibitem{KZR+94}
M.~A. Krzyzowski {\it et~al.}, Phys. Rev. B {\bf 50},  18505  (1994).

\bibitem{MP80}
V.~I. Marchenko and A.~Y. Parshin, Sov. Phys. JETP {\bf 52},  129  (1980).
  

\bibitem{Noz92}
P. Nozi\`{e}res,  in {\em Solids Far from Equilibrium}, edited by C.
  Godr\`{e}che (Cambridge University Press, Cambridge, 1992).

\bibitem{Lap94}
J. Lapujoulade, Surf. Sci. Rep. {\bf 20},  191  (1994).

\bibitem{SPG94}
J. Stewart, O. Pohland, and J.~M. Gibson, Phys. Rev. B {\bf 49},  13848
  (1994).

\bibitem{SMRBH94}
S. Speller {\it et~al.}, Surf. Sci. Lett. {\bf 312},  L748  (1994).

\bibitem{AIP}
 in {\em American Institute of Physics Handbook}, edited by D.~G. et~al. (Mc
  Graw-Hill, New York, 1972).

\bibitem{SNS94}
S. Swaminarayan, R. Najafabadi, and D.~J. Srolovitz, Surf. Sci. {\bf 306},  367
   (1994).

\end{thebibliography}

\end{document}